\documentclass[12pt]{article}
\usepackage[compress]{cite}
\usepackage{hyperref}
\hypersetup{colorlinks,urlcolor=blue,citecolor=blue,linkcolor=blue,filecolor=black}
\usepackage{graphicx}
\usepackage{amsmath}
\usepackage{bm}
\usepackage{amssymb}
\usepackage{footmisc}
\usepackage{slashed}

\newcommand{\z}[1]{\zeta_{#1}}

\newcommand{\cN}{\mathcal{N}}
\newcommand{\bra}[1]{\langle#1\mid}
\newcommand{\ket}[1]{\mid#1\rangle}

\newcommand{\MS}{{\mathrm{MS}}}

\def\Z{\mathcal{Z}}
\def\D{\mathcal{D}}
\def\N{\mathcal{N}}

\begin{document}

\title{Anomalous dimensions of twist-two operators in extended $\mathcal{N}=2$ and $\mathcal{N}=4$ super Yang-Mills theories}

\author{\sc B.A. Kniehl$^{(a)}$, 
V.N. Velizhanin$^{(a),(b)}$
\\[3mm]
\it (a) II. Institut f\"ur Theoretische Physik, Universit\"at Hamburg,\\
\it Luruper Chaussee 149, 22761 Hamburg, Germany
\\[2mm]
\it (b) Theoretical Physics Division, NRC ``Kurchatov Institute'',\\
\it Petersburg Nuclear Physics Institute,\\
\it Orlova Roscha, Gatchina, 188300 St.~Petersburg, Russia
}


\date{}

\maketitle

\begin{abstract}
  We perform direct diagrammatic calculations of the anomalous dimensions of twist-two operators in extended $\mathcal{N}=2$ and $\mathcal{N}=4$ super Yang-Mills theories (SYM). In the case of $\mathcal{N}=4$ SYM, we compute the four-loop anomalous dimension of the twist-two operator for several fixed values of Lorentz spin. This is the first direct diagrammatic calculation of this kind, and we confirm results previously obtained by means of integrability. For $\mathcal{N}=2$ SYM, we obtain the general result for the anomalous dimension at third order of perturbation theory and find the three-loop Cusp anomalous dimension.
\end{abstract}

\newpage

\section{Introduction}

Twist-two operators play an important role in quantum field theory. First of all, they appear in the study of deep-inelastic scattering processes in Quantum Chromodynamics (QCD), where they give the leading contribution to the electron-proton cross section through the operator product expansion. The anomalous dimensions of twist-two operators enter the Dokshitzer-Gribov-Lipatov-Altarelli-Parisi (DGLAP)~\cite{Gribov:1972ri,Gribov:1972rt,Altarelli:1977zs,Dokshitzer:1977sg} evolution equation for the parton distribution functions with energy scale, which are widely used for the interpretation of experimental data. The leading-order anomalous dimensions of twist-two operators were computed fifty years ago by Gross and Wilczek~\cite{Gross:1973ju,Gross:1974cs}. While the second-order results were obtained just a few years later~\cite{Floratos:1977au,GonzalezArroyo:1979df,Floratos:1978ny,GonzalezArroyo:1979he,Gonzalez-Arroyo:1979kjx,Curci:1980uw,Furmanski:1980cm}, it took more than twenty years until the third-order results became available~\cite{Larin:1996wd,Moch:2004pa,Vogt:2004mw}. Results of higher orders are available only for fixed values $M$ of Lorentz spin~\cite{Velizhanin:2011es,Velizhanin:2014fua,Moch:2017uml,Moch:2018wjh,Herzog:2018kwj,Moch:2023tdj}.

For the interpretation of experimental data, it is necessary to know the
anomalous dimensions of the twist-two operators for arbitrary value of $M$.
In fact, the evolution of the parton distribution functions, measured at fixed energy, is usually performed in Bjorken $x$ space, and the evolution kernels in $x$ space emerge from the anomalous dimensions in $M$ space via inverse Mellin transformation.

The other great interest in twist-two operators arises from the study of the
anti-de~Sitter/conformal field theory (AdS/CFT) correspondence~\cite{Maldacena:1997re,Gubser:1998bc,Witten:1998qj}. It was discovered~\cite{Minahan:2002ve} that integrability can aid with the computations of anomalous dimension of composite operators. Namely, the asymptotic Bethe ansatz was proposed as a solution to the spectral problem~\cite{Minahan:2002ve,Beisert:2003tq,Beisert:2003yb,Bena:2003wd,Kazakov:2004qf,Beisert:2004hm,Arutyunov:2004vx,Staudacher:2004tk,Beisert:2005fw,Eden:2006rx,Beisert:2006ez,Beisert:2010jr}, including twist-two operators~\cite{Staudacher:2004tk} that belong to the $\mathfrak{sl}(2)$ closed subsector of the full group symmetry in $\cN=4$ SYM. Later, integrability was extended to the finite length operators~\cite{Arutyunov:2009zu,Gromov:2009tv,Arutyunov:2009ur,Bombardelli:2009ns,Gromov:2009bc,Arutyunov:2009ax,Gromov:2013pga,Gromov:2014caa} and also to deformed $\cN=4$ SYM~\cite{Berenstein:2004ys,Beisert:2005if} and to two-loop $\cN=2$ SYM~\cite{Belitsky:2004sc,Belitsky:2005bu}.

In the context of integrability, the following scalar twist-two operator is usually considered:
\begin{equation}
{\mathcal{O}}^M_{\Z\Z}
=\Z\D_{\mu_1}\cdots\D_{\mu_{{}M}}\Z\,,
\label{ZZ}
\end{equation}
where $\Z$ is a complex scalar field and $\D_\mu$ is the covariant derivative.\footnote{%
In $\mathcal{N}=4$ SYM, there are three complex scalar fields, $\Phi_j=A_j+iB_j$, where the index $j=1,2,3$ refers to the internal symmetry group. Specifically, we denote $\Z=\Phi_3$, $A=A_3$, and $B=B_3$.} 
Its advantage is that it does not mix with other operators under renormalization because it belongs to the closed $\mathfrak{sl}(2)$ subsector \cite{Beisert:2003jj}. On the other hand, such an operator is not considered in QCD, where scalar fields are absent.

In this paper, we consider the anomalous dimension of a twist-two operator that is closely related to $\mathcal{O}_{\Z\Z}^M$. For $\cN=4$ SYM, we compute the first three even Mellin moments at four loops and confirm, for the first time by direct diagrammatic calculation, the correctness of results obtained with the help of integrability.
For $\cN=2$ SYM, we perform calculations through three-loop order for a rather wide range of  $M$ values and find the general form of the three-loop anomalous dimension, for arbitrary value of $M$, using modern tools of number theory. From the general result thus obtained, we find the analytic result for the three-loop Cusp anomalous dimension.

\section{Calculation}

The twist-two operator usually studied in the context of integrability, introduced in Eq.~\eqref{ZZ}, involves two appearances of the complex scalar field $\Z=A+i B$. As we use the Feynman rules from Refs.~\cite{Gliozzi:1976qd,Avdeev:1980bh}, we choose to work with the two real scalar fields, $A$ and $B$, instead. Therefore, our calculations refer to the following scalar twist-two operator:
\begin{equation}
{\mathcal{O}}^M_{AB}
=A\D_{\mu_1}\cdots\D_{\mu_{{}M}}B\,.\label{AB}
\end{equation}
Notice that the operators ${\mathcal{O}}^M_{\Z\Z}$ and ${\mathcal{O}}^M_{AB}$ share the same anomalous dimension.

For the calculation of the anomalous dimension, we consider the one-particle-irreducible Green's function that is obtained by contracting the matrix element of the local operator ${\mathcal{O}_{AB}^M}$ with the source term $J_M = \Delta^{\mu_1} \ldots \Delta^{\mu_M}$ as
\begin{equation}
\label{GijTRNS}
\Gamma_{\mathcal{O}_{AB}^M}=
J_M\bra{A(p)}{\mathcal O}_{AB}^{M}\ket{B(p)}\,,
\end{equation}
where $p$ denotes the momentum of the external scalar lines and $\Delta$ is a light-like vector, with $\Delta^2 = 0$.
The contraction with the source term $J_M$ allows us to write a general expression for the corresponding projector, which may be found similarly as in Ref.~\cite{Bierenbaum:2009mv}. In our case, the projector is just a fully symmetric, traceless tensor.

We calculate the renormalization constants as in Ref.~\cite{Collins:1974bg} (see also Refs.~\cite{Tarasov:1976ef,Vladimirov:1979zm,Tarasov:1980au,Larin:1993tp}) by making use of multiplicative renormalizability of Green's functions.
The renormalization constants $Z_\Gamma$ relate the dimensionally regularized one-particle-irreducible Green's functions with the renormalized ones as
\begin{equation}
\label{multren}
\Gamma_{\mathrm{renormalized}}\left(\frac{Q^2}{\mu^2},\alpha,g^2\right)=\lim_{\epsilon \rightarrow 0}
Z_\Gamma\left(\frac{1}{\epsilon},\alpha,g^2\right)
\Gamma_{\mathrm{bare}}\left(Q^2,\alpha_{\mathrm{B}},g^2_\mathrm{B},\epsilon\right)\,,
\end{equation}
where $Q^2=p^2$, $g^2$ is the charge, $\alpha$ is the gauge fixing parameter, $\mu$ is the 't~Hooft mass, $4-2\epsilon$ is the space-time dimension, and bare quantities are labeled by the subscript $\mathrm{B}$.
The latter are related to their renormalized counterparts as
\begin{equation}\label{gbex}
g^2_{\mathrm{B}}=
\mu^{2\epsilon}\left[g^2+\sum_{n=1}^{\infty}a^{(n)}\left(g^2\right)\epsilon^{-n}\right]\,,
\qquad\alpha_{\mathrm{B}}=\alpha Z_v\,.
\end{equation}
To obtain the anomalous dimension of the operator $\mathcal{O}^M_{AB}$, we subtract the renormalization constants of the external scalar fields, $Z_A$ and $Z_B$, as
\begin{equation}
Z_{\mathcal{O}^{\,M}_{AB}}=Z_A^{-1/2}Z_B^{-1/2}\,Z_{\Gamma}\,.\label{gbz}
\end{equation}
We have $Z_A=Z_B$.
The three-loop results for $Z_v$ and $Z_A$ in $\mathcal{N}=4$ and $\mathcal{N}=2$ SYM were obtained in Refs.~\cite{Avdeev:1981ew,Velizhanin:2008rw}. We refrain from listing them here, but refer instead to Eqs.~(10), (12), (14), and (16) in Ref.~\cite{Velizhanin:2008rw}, where $Z_v$ and $Z_A$ are called $Z_g$ and $Z_s$, respectively.

Renormalization constants within $\MS$-like schemes do not depend on dimensionful parameters such as masses and momenta~\cite{Collins:1974bg} and have the following structure:
\begin{equation}
\label{eq:5}
Z_\Gamma\!\left(\frac{1}{\epsilon},\alpha,g^2\right)=
1+\sum^\infty_{n=1}c_\Gamma^{(n)}\!\left(\alpha,g^2\right)\epsilon^{-n}\,.
\end{equation}
The respective anomalous dimensions are defined in terms of the renormalization constants as
\begin{equation}
\label{defga}
\gamma_\Gamma(\alpha,g^2)=
g^2\frac{\partial}{\partial g^2}\ c^{(1)}_\Gamma(\alpha,g^2)=\sum^\infty_{n=1}\gamma_\Gamma^{(n-1)}g^{2n}\,,
  \qquad g^2=\frac{\lambda}{16\pi^2}\,,
\end{equation}
where $\lambda=g^2_{\mathrm {YM}}N_c$ is the 't~Hooft coupling constant.

We use the program package \texttt{DIANA}~\cite{Tentyukov:1999is}, which calls the package \texttt{QGRAF}~\cite{Nogueira:1991ex}, to generate all the diagrams.
The Feynman rules for the vertices of the scalar operator ${\mathcal{O}}_{AB}$ for even $M$ are related to the quark operator in QCD for odd $M$,\footnote{%
Notice that one Lorentz index of the quark operator is related to the $\gamma$ matrix, as may be observed, for example, from Ref.~\cite{Bierenbaum:2009mv}.} and the explicit form of the operator vertices contributing through the four-loop order are listed in the Appendix. A custom-made code in \texttt{FORM} \cite{Kuipers:2012rf} language produces the initial structures for these operator vertices, from which the residual terms emerge as permutations. The projectors for the Feynman diagrams involving operators are symmetric traceless structures, which may be generated using \texttt{FORM} in a rather simple way. As the operators and projectors are fully symmetric under Lorentz transformations, we can symmetrize either operators or projectors, and we choose the latter option.
We compute for each diagram the traces of the $\gamma$ matrices and simplify the structures related to the $\alpha_i$ and $\beta_i$ matrices of the Feynman rules for Yukawa vertices by adopting from Ref.~\cite{Avdeev:1980bh} simple relationships between them.
We use the \texttt{FORM} package \texttt{COLOR}~\cite{vanRitbergen:1998pn} to evaluate the color traces.
For the identification of the momenta inside topologies, we use a custom-made code written in \texttt{MATHEMATICA} language.

Because the right-hand side of Eq.~(\ref{multren}) contains the bare gauge fixing parameter $\alpha_{\mathrm{B}}$, all calculations 
should a priori be performed for generic value of $\alpha$,
i.e.\ using the gluon propagator in the form $[g_{\mu\nu}-(1-\alpha)q_\mu q_\nu/q^2]/q^2$. 
However, if we are interested in the result at $\ell$-th order of perturbation theory, we are entitled to perform the calculation at this order in Feynman gauge, with $\alpha=1$, and to allow for $\alpha$ to be variable only at the lower orders.
To obtain the result for the renormalization constant from Eq.~(\ref{multren}), we put $\alpha=1$ only after expansion of the right-hand side of Eq.~(\ref{multren}).

We adopt the dimensional-reduction ($\overline{\mathrm{DR}}$) scheme~\cite{Siegel:1979wq}, which restores supersymmetry within dimensional regularization.
Inside the \texttt{FORM} environment, the $\overline{\mathrm{DR}}$ scheme can be implemented by properly using the \texttt{trace} statement for manipulations of the Dirac algebra. In particular, use of the \texttt{trace4} operation, which assumes that the $\gamma$ matrices are defined in four dimensions, with option \texttt{notrick}, which ensures that specifically four-dimensional relations are avoided in the final stage of trace taking, is found to yield correct results through four loops in $\mathcal{N}=4$ SYM.

At the end of these procedures, the intermediate results are expressed in terms of massless propagator-type integrals and collected in files of typically several gigabytes. At the four-loop level, we use the \texttt{FORM} program \texttt{FORCER} \cite{Ruijl:2017cxj} for the parametric reduction of these integrals to master integrals. This requires high-performance computers, with large amounts of RAM and fast drives with large capacities for the temporary files.

\boldmath
\section{Three-loop anomalous dimension in $\cN=2$ SYM}
\unboldmath

At one loop, we have only two Feynman diagrams with operator insertions, and these yield the same result both in $\cN=2$ and $\cN=4$ SYM. The difference comes from the renormalization of the external legs, i.e.\ from the anomalous dimension of the scalar field. In $\cN=2$ SYM, the one-loop anomalous dimension of $\mathcal{O}_{AB}^M$ reads
\begin{equation}
  \gamma_{AB}^{(0)}(M)=8S_1(M)-4\,,
\label{ADN2L1}
\end{equation}
with harmonic sum
\begin{equation}
S_1(M)=\sum_{j=1}^M\frac{1}{j}\,,
\end{equation}
in agreement with Ref.~\cite{Belitsky:2005bu}.
The respective result in $\cN=4$ SYM emerges from the right-hand side of Eq.~\eqref{ADN2L1} by omitting the last constant.

At two loops, there are 79 nonvanishing diagrams, and the result in $\cN=2$ SYM can be written as
\begin{equation}
\gamma_{AB}^{(1)}(M)=16(S_{-3}+S_3-2S_{1,-2}-2S_{1,2}-2S_{2,1})
+32 S_1\,,
\label{ADN2L2}
\end{equation}
with nested harmonic sums, recursively defined as \cite{Vermaseren:1998uu,Blumlein:1998if}
\begin{eqnarray}
S_a (M)=\sum^{M}_{j=1} \frac{(\mbox{sgn}(a))^{j}}{j^{\vert a\vert}}\,,\qquad
S_{a_1,\ldots,a_n}(M)=\sum^{M}_{j=1} \frac{(\mbox{sgn}(a_1))^{j}}{j^{\vert a_1\vert}}
\,S_{a_2,\ldots,a_n}(j)\, ,\label{vhs}
\end{eqnarray}
in agreement with Ref.~\cite{Belitsky:2005bu}.
Again, the last term in Eq.~\eqref{ADN2L2} is absent in $\cN=4$ SYM.

For the calculation of the three-loop anomalous dimension in $\N=2$ SYM, we exploit the results for the anomalous dimension in $\N=4$ SYM, which is known through seven-loop order for arbitrary value of $M$. For this purpose, we calculate the Feynman diagrams for arbitrary numbers of scalar and fermion fields, $N_s$ and $N_f$. The total number of three-loop diagrams is 3593, while the number of diagrams proportional $N_s$ or/and $N_f$ is 1634. Correspondingly, the three-loop anomalous dimension can be separated into two parts, as
\begin{equation}
\gamma_{AB}=\hat{\gamma}_{AB}+\gamma_{AB}^{N_s,N_f}\,.
\end{equation}
For $\N=2$, this is
\begin{equation}
\gamma_{AB}^{\N=2}=\hat{\gamma}_{AB}+\gamma_{AB}^{N_s=1,N_f=2}\,,
\end{equation}
while, for $\N=4$, this is
\begin{equation}
  \gamma_{AB}^{\N=4}=\hat{\gamma}_{AB}+\gamma_{AB}^{N_s=3,N_f=4}\,.
\end{equation}
As $\gamma_{AB}^{\N=4}$ is known from integrability, we can find $\gamma_{AB}^{\N=2}$ as
\begin{equation}
  \gamma_{AB}^{\N=2}=\gamma_{AB}^{\N=4}-\hat{\gamma}_{AB}^{\cN=4-\cN=2}\,,\qquad
  \hat{\gamma}_{AB}^{\cN=4-\cN=2}=\gamma_{AB}^{N_s=3,N_f=4}-\gamma_{AB}^{N_s=1,N_f=2}\,.
\label{eq:gammaab}
\end{equation}
It is important to observe that the Feynman diagrams that contain the operator vertex with three gauge fields do not contribute to $\gamma_{AB}^{N_s,N_f}$. This operator, being contracted with the projector, has three open Lorentz indices, and the number of terms in the corresponding expression grows very fast with $M$. The procedure described above makes it possible to exclude this operator from consideration. In this way, we may obtain results through rather large Lorentz spin, through $M=22$. Our results for the difference $\hat{\gamma}_{AB}^{\cN=4-\cN=2}$ for $M=2,\ldots,22$ read:
\begin{eqnarray}
&&\Big\{
336,
\frac{8725}{18},
\frac{71687}{125},
\frac{23551099199}{37044000},
\frac{6834557182739}{10001880000},
\frac{16678419497792}{23111983125},\nonumber\\
&&\quad
\frac{12857387174422861663}{17061081047010000},
\frac{51166461615253770629}{65514551220518400},
\frac{518156867201499886794349}{643745980292813798400},\nonumber\\
&&\quad
\frac{3647666896520541464030933897}{4415453678828409843225600},
\frac{41048732781908245232317902859}{48569990467112508275481600}
\Big\}\,.\label{resdiffN4N2}
\end{eqnarray}

To find the general result for the three-loop anomalous dimension of the operator $\mathcal{O}_{AB}^M$ in $\cN=2$ SYM for arbitrary value of $M$, we first reconstruct the general form of the difference $\hat{\gamma}_{AB}^{\cN=4-\cN=2}$ defined in Eq.~\eqref{eq:gammaab}.
For the reconstruction, we use the same method as in our related previous computations~\cite{Velizhanin:2010cm,Velizhanin:2013vla,Marboe:2014sya,Marboe:2016igj,Kniehl:2020rip,Kniehl:2021ysp}, i.e.\ we write down the most common basis, which consists of nested harmonic sums, and fix the coefficients in front of the latter using advanced tools of number theory, including the \texttt{fplll} \cite{fplll} implementation of the Lenstra-Lenstra-Lovasz algorithm \cite{Lenstra82factoringpolynomials}, which allows us to reduce the matrix obtained from the system of Diophantine equations to a form in which the rows are the solutions of the system with the minimal Euclidean norm.
At three loops, we initially expect for the difference $\hat{\gamma}_{AB}^{\cN=4-\cN=2}$ that the basis includes the harmonic sums through weight $4$, as the three-loop anomalous dimension in $\cN=4$ SYM contains the harmonic sums through weight $5$. However, such a simple assumption regarding the basis does not provide any reasonable result, and we are led to include also terms with the nested harmonic sums being divided by $M$ and $(M+1)$. The basis thus extended includes 49 terms altogether. Injecting the first ten numbers from Eq.~\eqref{resdiffN4N2}, for $M=2,\ldots,20$, we then find
\begin{eqnarray}
\hat{\gamma}_{AB}^{\cN=4-\cN=2}&=&
{}-32(S_{3,1}+S_{1,3}-2 S_{-2,-2})
-128(S_{-3}+S_3-2 S_{1,-2}-2 S_{1,2}-2 S_{2,1})
\nonumber\\[3mm]
&&{}+64\frac{S_1 S_{-2}}{M(M+1)}\,.\label{N2l3}
\end{eqnarray}
Inserting $M=22$ in Eq.~\eqref{N2l3}, we recover the last number in Eq.~\eqref{resdiffN4N2}, which nicely confirms the validity of the all-$M$ formula in Eq.~\eqref{N2l3}.

Notice that the last term in Eq.~\eqref{N2l3} exactly coincides with the result for the wrapping correction for the anomalous dimension,
$\gamma_{\Z\Z}^{(2),\,\beta\text{-def},\,\N=4,\,\text{wrap}}$, of the twist-two operator in $\beta$ deformed $\cN=4$ SYM \cite{deLeeuw:2010ed}.
The above result for the three-loop anomalous dimension in $\N=2$ SYM can thus be rewritten in the following form
\begin{eqnarray}
\gamma_{AB}^{(2),\,\N=2}=\gamma_{\Z\Z}^{(2),\,\N=4}
+8\gamma_{\Z\Z}^{(1),\,\N=4}
+\gamma_{\Z\Z}^{(2),\,\beta\text{-def},\,\N=4,\,\text{wrap}}
+32(S_{3,1}+S_{1,3}-2S_{-2,-2})\,,\label{N2l3rN4}
\end{eqnarray}
where $\gamma_{\Z\Z}^{(\ell-1),\,\N=4}$ is the $\ell$-loop anomalous dimension in $\N=4$ SYM and the last term is the additional contribution that appears in $\N=2$ SYM. The first three terms are amenable to integrability, while the last term is new and violates the simple structure of results that can be computed by integrability. It will be very interesting to try and understand its relation with integrability, too.

\boldmath
\subsection{Large-$M$ limit}
\unboldmath

Using the general results for the three-loop anomalous dimension of the twist-two operator $\mathcal{O}_{AB}^M$ in $\N=2$ SYM from Eqs.~\eqref{eq:gammaab} and \eqref{N2l3}, we find the following large-$M$ asymptotic behavior:
\begin{equation}
\gamma_{M\to\infty}^{\cN=2}=
\gamma_{M\to\infty}^{\cN=4}
+\left(\gamma_{\mathrm{Cusp}}^{\cN=2}-\gamma_{\mathrm{Cusp}}^{\cN=4}\right)\ln M
-4g^2+0\times g^4+4g^6(48\z3+5\z4)\,,\label{SFfN2}
\end{equation}
where $\zeta_n=\zeta(n)$ is Riemann's zeta function and the three-loop Cusp anomalous dimension in $\cN=2$ SYM reads
\begin{equation}
\gamma_{\mathrm{Cusp}}^{\cN=2}=\gamma_{\mathrm{Cusp}}^{\cN=4}
+0\times g^2+32g^4+32g^6\,(4\z2-\z3)\,.\label{CuspN2}
\end{equation}
The corresponding three-loop results in $\cN=4$ SYM read \cite{Kotikov:2004er}
\begin{eqnarray}
\gamma_{M\to\infty}^{\cN=4}&=&
\gamma_{\mathrm{Cusp}}^{\cN=4}\ln M
+0\times g^2
-24\z3g^4
+32g^6 (\z2 \z3 + 5\z5)\,,
\nonumber\\
\gamma_{\mathrm{Cusp}}^{\cN=4}&=&
8g^2
-16\z2g^4
+176\z4g^6\,.\label{CuspN4}
\end{eqnarray}
The one- and two-loop results in Eq.~\eqref{CuspN2} have originally been obtained in Refs.~\cite{Gorsky:2002ju,Belitsky:2003ys,Belitsky:2005bu}.

It will be interesting to compute $\gamma_{\mathrm{Cusp}}^{\cN=2}$ with the methods that were used for the calculations of the four-loop Cusp anomalous dimensions in QCD and $\N=4$ SYM~\cite{Lee:2016ixa,Lee:2019zop,Henn:2019swt,Huber:2019fxe,vonManteuffel:2020vjv}.

\section{Four-loop anomalous dimension in $\cN=4$ SYM}
\unboldmath

The initial number of four-loop diagrams is 179977, but 33721 of them vanish due to zero color factor, so that we are left with 146256 diagrams.
Their evaluation with \texttt{FORCER} demands several hours of computer time for $M=2$ and several days for $M=6$. Finally, using Eq.~\eqref{gbz}, we obtain
\begin{eqnarray}
\gamma_{AB}
(2)&=&
12 g^2-48 g^4+336 g^6+g^8 \left(-2496+576 \z3-1440 \z5- 360 \z5 \frac{48}{N_c^2}\right)\,,\nonumber\\
\gamma_{AB}(4)&=&
\frac{50}{3}g^2
-\frac{1850 }{27}g^4
+\frac{241325 }{486}g^6\nonumber\\&&
{}+g^8 \left[-\frac{8045275}{2187}+\frac{114500}{81} \z3-\frac{25000}{9} \z5+\frac{25}{9}(21 + 70  \z3 - 250 \z5)  \frac{48}{N_c^2}\right]\,,\nonumber\\
\gamma_{AB}(6)&=&
\frac{98 }{5}g^2
-\frac{91238 }{1125}g^4
+\frac{300642097 }{506250}g^6
+g^8 \left[-\frac{393946504469}{91125000}\right.\nonumber\\&&
{}+\left.\frac{11736088 }{5625}\z3
-\frac{19208 }{5}\z5+\frac{49}{600}(1357 + 4340 \z3 - 11760 \z5) \frac{48}{N_c^2}\right]\,.
\end{eqnarray}
The planar parts of the above results coincide with those obtained with the help of integrability~\cite{Gromov:2009tv,Staudacher:2004tk,Bajnok:2008qj,Marboe:2014gma}.
The nonplanar contributions were calculated by direct diagrammatic methods in Refs.~\cite{Velizhanin:2009gv,Velizhanin:2010ey,Velizhanin:2014zla,Kniehl:2020rip,Kniehl:2021ysp} and by applying the method of asymptotic expansions to the four-point functions of length-two half--Bogomol'nyi-Prasad-Sommerfield operators~\cite{Fleury:2019ydf}.

Unfortunately, the technique successfully applied in $\cN=4$ SYM above fails for $\cN=2$ SYM due an inadequacy of the $\overline{\mathrm{DR}}$ scheme. For example, even for the renormalization constant of the scalar fields, we obtain a non-zero second pole for the non-planar part of the four-loop result in $\cN=2$ SYM, although all higher poles should be absent for this part because it appears for the first time at four loops, while higher poles are related to lower loops. The failure of the $\overline{\mathrm{DR}}$ scheme in $\cN=2$ SYM has also been encountered at the three-loop order~\cite{Avdeev:1981ew,Velizhanin:2008rw}. Specifically, the three-loop $\beta$ function of the Yukawa coupling has been found to be non-zero, in conflict with a well-established theorem \cite{Novikov:1983uc,Novikov:1985rd}. For a critical scrutiny of the calculational capabilities of the
$\overline{\mathrm{DR}}$ scheme in $\mathcal{N}=1,2,4$ SYM, we refer to Ref.~\cite{Avdeev:1982xy}. This problem deserves a more detailed study, which, however, reaches beyond the scope of this paper and is left for future work.

\section{Conclusions}

We performed direct diagrammatic calculations of the anomalous dimensions of the twist-two operators $\mathcal{O}_{AB}^M$ in extended $\cN=2$ and $\cN=4$ SYM. We confirmed the results obtained with the help of integrability in $\cN=4$ SYM at forth order of perturbation theory. 
In $\cN=2$ SYM, we found the general expression for the three-loop anomalous dimension of the twist-two operator $\mathcal{O}_{AB}^M$ for arbitrary Lorentz spin $M$, which is given by Eqs.~\eqref{eq:gammaab} and \eqref{N2l3}.
In contrast to the situation at one and two loops, this result can only partly by expressed in terms of anomalous dimensions that can be computed with the help of integrability, as in Eq.~\eqref{N2l3rN4}.
The residual contribution in Eq.~\eqref{N2l3rN4} has a very simple and uniform structure, and it would be interesting to understand its relation to integrability.
Furthermore, from the all-$M$ expression in Eqs.~\eqref{eq:gammaab} and \eqref{N2l3}, we obtained, for the first time, the three-loop Cusp anomalous dimension of $\cN=2$ SYM, which is presented in Eq.~\eqref{CuspN2}.
It will be interesting to compute $\gamma_{\mathrm{Cusp}}^{\cN=2}$ with the methods that were used for the calculations of the four-loop Cusp anomalous dimensions in QCD and $\N=4$ SYM~\cite{Lee:2016ixa,Lee:2019zop,Henn:2019swt,Huber:2019fxe,vonManteuffel:2020vjv}.


 \subsection*{Acknowledgments}

The work of B.A.K. was supported by the German Research Foundation DFG through Grant No.~KN~365/16-1.
The work of V.N.V. was supported by the Russian Science Foundation under Grant No.~23-22-00311.

\section*{Appendix}

In this Appendix, we list the Feynman rules for all operator vertices that can appear in calculations of four-loop anomalous dimensions in extended supersymetric Yang-Mills theory.
All momenta are taken to be incoming to the vertex, adding up to zero, $\Delta$ is some light-like vector, and the color indices of gluon lines are denoted by lowercase Latin letters.

\allowdisplaybreaks

\begin{eqnarray}
&&\hspace*{-10mm}
\scalebox{1}[1]{\includegraphics[trim = 0mm 0mm 0mm 0mm, clip,height=2cm,angle=0]{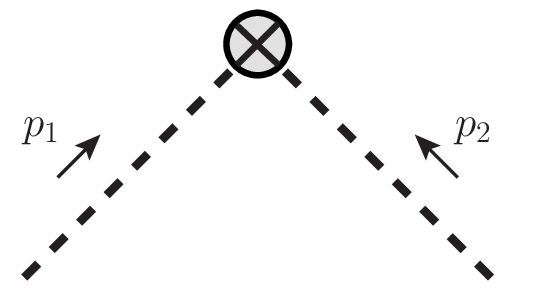}}\nonumber\\[-15mm]
&&\hspace*{35mm}
\mathcal{O}^{V(0)}=
(\Delta\cdot p_1)^N\,,
\label{OpVer0}\\[10mm]
%
&&\hspace*{-10mm}
\scalebox{1}[1]{\includegraphics[trim = 0mm 6mm 0mm 0mm, clip,height=2.24cm,angle=0]{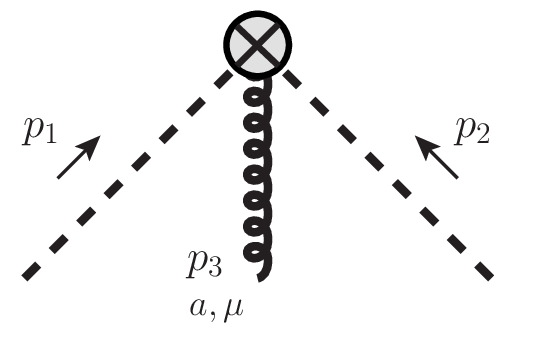}}\nonumber\\[-15mm]
&&\hspace*{35mm}
\mathcal{O}^{V(1)}=\mathcal{O}^{p_3}_{\,t^a}=
\sum_{j=0}^{N-1}(-\Delta\cdot p_2)^j (\Delta\cdot p_1)^{N-j-1}\,,
\label{OpVer1}\\[5mm]
%
%
&&\hspace*{-10mm}
\scalebox{1}[1]{\includegraphics[trim = 0mm 6mm 0mm 0mm, clip,height=2.24cm,angle=0]{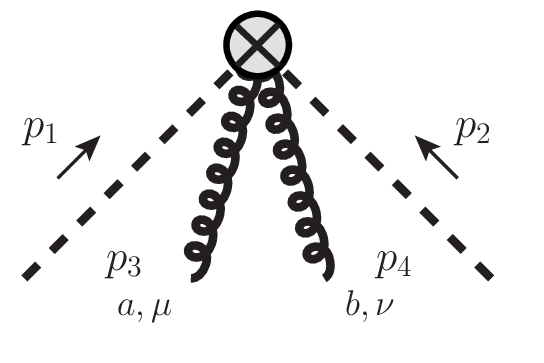}}\nonumber\\[-15mm]
&&\hspace*{35mm}
\mathcal{O}^{p_3 \,p_4}_{\,t^a\;t^b}=
\sum_{j=0}^{N-2}\sum_{k=j+1}^{N-1}(-\Delta\cdot p_2)^j (\Delta\cdot p_1)^{N-k-1}\times\nonumber\\
&&\hspace*{75mm}\Big[
(t^a t^b)(\Delta p_1+\Delta p_3)^{k-j-1}
\Big]\,,\nonumber\\[3mm]
&&\hspace*{40mm}
\mathcal{O}^{V(2)}=\mathcal{O}^{p_3 \,p_4}_{\,t^a\;t^b}+\mathcal{O}^{p_4 \,p_3}_{\,t^b\;t^a}\,,
\label{OpVer2}\\[5mm]
%
&&\hspace*{-10mm}
\scalebox{1}[1]{\includegraphics[trim = 0mm 6mm 0mm 0mm, clip,height=2.24cm,angle=0]{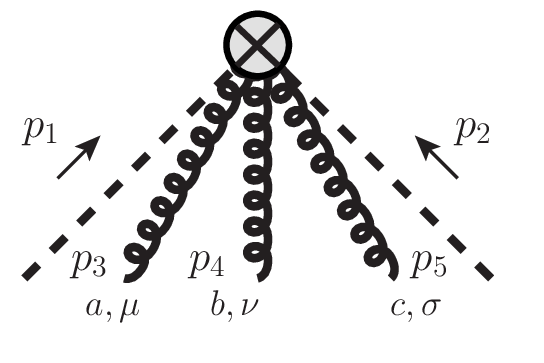}}\nonumber\\[-15mm]
&&\hspace*{35mm}
\mathcal{O}^{p_3 \,p_4\,p_5}_{\,t^a\;t^b\;t^c}=
\sum_{j=0}^{N-3}\sum_{k=j+1}^{N-2}\sum_{l=k+1}^{N-1}(-\Delta\cdot p_2)^j (\Delta\cdot p_1)^{N-l-1}
\times\nonumber\\
&&\hspace*{40mm}\Big[
(t^a t^b t^c)(\Delta p_1+\Delta p_3+\Delta p_4)^{k-j-1}(\Delta p_1+\Delta p_3)^{l-k-1}
\Big]\,,\nonumber\\[3mm]
&&\hspace*{0mm}
\mathcal{O}^{V(3)}=\mathcal{O}^{p_3 \,p_4\,p_5}_{\,t^a\;t^b\;t^c}
+\mathcal{O}^{p_3 \,p_5\,p_4}_{\,t^a\;t^c\;t^b}
+\mathcal{O}^{p_4 \,p_3\,p_5}_{\,t^b\;t^a\;t^c}
+\mathcal{O}^{p_4 \,p_5\,p_3}_{\,t^b\;t^c\;t^a}
+\mathcal{O}^{p_5 \,p_3\,p_4}_{\,t^c\;t^a\;t^b}
+\mathcal{O}^{p_5 \,p_4\,p_3}_{\,t^c\;t^b\;t^a}\,,
\label{OpVer3}\\[5mm]
%
&&\hspace*{-10mm}
\scalebox{1}[1]{\includegraphics[trim = 0mm 6mm 0mm 0mm, clip,height=2.24cm,angle=0]{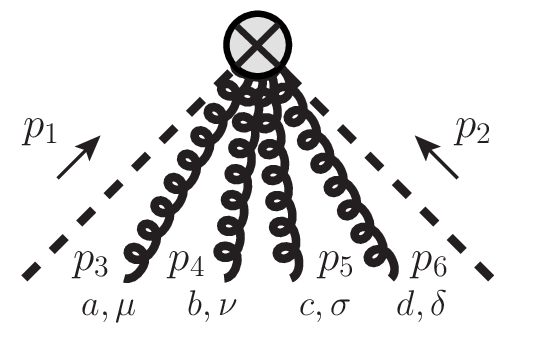}}\nonumber\\[-15mm]
&&\hspace*{35mm}\mathcal{O}^{p_3 \,p_4\,p_5\,p_6}_{\,t^a\;t^b\;t^c\;t^d}=
\sum_{j=0}^{N-4}\sum_{k=j+1}^{N-3}\sum_{l=k+1}^{N-2}\sum_{m=l+1}^{N-1}(-\Delta\cdot p_2)^j (\Delta\cdot p_1)^{N-m-1}\times\nonumber\\
&&\hspace*{-5mm}\Big[
(t^a t^b t^c t^d)(\Delta p_1+\Delta p_3+\Delta p_4+\Delta p_5)^{k-j-1}(\Delta p_1+\Delta p_3+\Delta p_4)^{l-k-1}(\Delta p_1+\Delta p_3)^{m-l-1}
\Big]\,,\nonumber\\[3mm]
&&\hspace*{0mm}
\mathcal{O}^{V(4)}=
\mathcal{O}^{p_3 \,p_4\,p_5\,p_6}_{\,t^a\;t^b\;t^c\;t^d}
+\mathcal{O}^{p_3 \,p_4\,p_6\,p_5}_{\,t^a\;t^b\;t^d\;t^c}
+\mathrm{20\ perm.}
+\mathcal{O}^{p_6 \,p_5\,p_3\,p_4}_{\,t^d\;t^c\;t^a\;t^b}
+\mathcal{O}^{p_6 \,p_5\,p_4\,p_3}_{\,t^d\;t^c\;t^b\;t^a}\,.
\label{OpVer4}
\end{eqnarray}

\end{document}